\documentclass[11pt]{article}
\usepackage{amssymb,latexsym,amsmath}
\usepackage{amsfonts, graphics}
\def\be{\begin{equation}}
\def\ee{\end{equation}}
\def\bea{\begin{eqnarray}}
\def\eea{\end{eqnarray}}
\def\beas{\begin{eqnarray*}}
\def\eeas{\end{eqnarray*}}

\newcommand{\R}{\mathbb R}

\def\open#1{\setbox0=\hbox{$#1$}
\baselineskip = 0pt
\vbox{\hbox{\hspace*{0.4 \wd0}\tiny $\circ$}\hbox{$#1$}}
\baselineskip = 11pt\!}
\def\fn{\open{f}\,}

\newcommand{\prfe}{\hspace*{\fill} $\Box$

\smallskip \noindent}
\newcommand{\prf}{{\bf Proof.\ }}

\def\be{\begin{equation}}
\def\ee{\end{equation}}
\def\bea{\begin{eqnarray}}
\def\eea{\end{eqnarray}}
\def\beas{\begin{eqnarray*}}
\def\eeas{\end{eqnarray*}}

\def\supp{\mathrm{supp}\,}

\begin{document}

\newtheorem{theorem}{Theorem}[section]
\renewcommand{\thetheorem}{\arabic{section}.\arabic{theorem}}
\newtheorem{definition}[theorem]{Definition}
\newtheorem{proposition}[theorem]{Proposition}
\newtheorem{example}[theorem]{Example}
\newtheorem{remark}[theorem]{Remark}
\newtheorem{cor}[theorem]{Corollary}
\newtheorem{lemma}[theorem]{Lemma}

\title{Formation of trapped surfaces for the
       spherically symmetric Einstein-Vlasov system}

\author{H{\aa}kan Andr\'{e}asson\\
        Mathematical Sciences\\
        Chalmers University of Technology\\
        G\"{o}teborg University\\
        S-41296 G\"oteborg, Sweden\\
        email: hand@math.chalmers.se\\
        \ \\
        Gerhard Rein\\
        Mathematisches Institut der Universit\"at Bayreuth\\
        D-95440 Bayreuth, Germany\\
        email: gerhard.rein@uni-bayreuth.de}

\maketitle

\begin{abstract}
We consider the spherically symmetric, asymptotically flat,
non-vacuum Einstein equations, using as matter model a collisionless
gas as described by the Vlasov equation.
We find explicit conditions on the initial data which guarantee
the formation of a trapped surface in the evolution which in particular
implies that weak cosmic censorship holds for these data.
We also analyze the evolution of solutions after a
trapped surface has formed and we show that the event horizon is
future complete. Furthermore we find that the apparent horizon
and the event horizon do not coincide. This behavior is analogous to
what is found in certain Vaidya spacetimes.
The analysis is carried out in Eddington-Finkelstein coordinates.
\end{abstract}
\section{Introduction}
\setcounter{equation}{0}
A major open problem in the mathematical analysis of
General Relativity are the geometric properties of spacetimes
in the case when singularities develop out
of regular initial data. Of particular interest is the
validity of the cosmic censorship conjecture which in
non-technical terms states that generically every
spacetime singularity which evolves out of regular, asymptotically
flat data is covered by an event horizon so that it cannot
be observed from infinity. For a precise form (or precise forms)
of a statement of cosmic censorship we refer to
\cite{Wald} or \cite{Chr99}. An important insight from
the mathematical analysis of this problem is that the answer
depends on the model used to describe the matter which collapses.
For dust which by definition is a perfect fluid with pressure
zero Christodoulou \cite{Chr84} showed that singularities which
violate cosmic censorship---so-called naked
singularities---arise for an open subset of spherically symmetric,
asymptotically flat data. Christodoulou also carried out an extensive analysis
of this problem with a massless scalar field as matter model
\cite{Chr91,Chr94,Chr99a}.
Again, there do exist spherically symmetric data which give
rise to naked singularities, but this behavior is shown to
be unstable, and for generic data the spacetime has a complete future null
infinity which is a more formal statement of the
validity of the weak cosmic censorship conjecture, cf.\ \cite{Chr99}.

In \cite{AKR2,AKR3,AR} gravitational collapse was studied using
as matter model a collisionless gas, i.e., the Vlasov equation.
The Einstein-Vlasov system describes an ensemble
of particles which interact only through gravity and where
collisions among the particles are neglected. It is a model
for astrophysical systems like galaxies or globular clusters,
although in astrophysics relativistic effects are most of the time
neglected and the non-relativistic limit
\cite{RR2, Rend94}
of the system, the Vlasov-Poisson
system, is used, cf.\ \cite{BT} and the references therein.
The main result for the Einstein-Vlasov system
in the cited investigations is that a class of spherically
symmetric, asymptotically flat initial data is identified
which lead to gravitational collapse and to a spacetime with a
complete future null infinity so that cosmic censorship
holds in the sense of \cite{Chr99}. The analysis is carried
out in Schwarzschild coordinates. It is shown that a
piece of the maximal development of the data is
covered by these coordinates which is sufficiently large to
deduce the desired conclusions, but the analysis is
hampered by the well known fact that Schwarzschild
coordinates cannot cover regions of spacetime which contain
trapped surfaces. Closely related to this defect
is the fact that these coordinates do not seem
to completely cover
the radially outgoing null geodesic
which generates the event horizon; at least
it has not been possible to show that they do.

It is hence natural to attempt an analogous analysis of
the Einstein-Vlasov system using coordinates
which do not break down at trapped surfaces. In the present
paper we propose to do so using Eddington-Finkelstein
coordinates. For the purpose of illustration we recall
that the Schwarzschild metric, when written in Schwarzschild
coordinates, takes the form
\[
ds^2 = -\left(1-\frac{2 M}{r}\right)\, dt^2 +
\left(1-\frac{2 M}{r}\right)^{-1}dr^2\,
+ r^2(d\theta^2 + \sin^2 \theta d\varphi^2).
\]
Here $M>0$ is a constant, the time coordinate $t\in \R$
coincides with the proper time of an observer who is at
rest at spatial infinity,  the area radius
$r>0$ labels the surfaces
of symmetry, i.e., the orbits of the group
$\mathrm{SO}(3)$ which have surface area $4 \pi r^2$
as measured in this metric, and
$\theta \in [0,\pi]$, $\varphi \in [0,2 \pi]$
parametrize these symmetry orbits.
The metric is non-singular with signature
$(-{} +{} +{} +{})$ in the region $r>2 M$, but
the coordinates break down at $r=2 M$.
In Eddington-Finkelstein coordinates $t$ is replaced by
\[
v := t + r + 2 M \ln (r-2 M).
\]
The Schwarzschild metric takes the form
\[
ds^2 = -\left(1-\frac{2 M}{r}\right)\, {dv}^2 + 2 dv\,dr\,
+ r^2(d\theta^2 + \sin^2 \theta d\varphi^2)
\]
which is completely regular for $r>0$, and $r=0$ is
revealed as the true spacetime singularity where
for example
\[
R^{\alpha \beta \gamma \delta} R_{\alpha \beta \gamma \delta} \sim r^{-6}
\ \mbox{as}\ r\to 0;
\]
$R_{\alpha \beta \gamma}^{\phantom{\alpha \beta \gamma}\delta}$
is the Riemann curvature tensor. The lines $v=const$ correspond to
radially ingoing null geodesics, and $v$ is referred to as
an advanced null coordinate or advanced time \cite[5.5]{HE}.
The line $r=2 M$ is also null and represents the event horizon,
while the surfaces of constant $v$ and $r$ are trapped if $r< 2 M$
in the sense that with increasing $v$ also the radially
outgoing null geodesics $dr/dv = 1-2 M/r$ at $r< 2 M$ move
towards the center.
We want to use such coordinates to study dynamic spacetimes
so we generalize the above form of the Schwarzschild metric
as follows:
\[
ds^2 = -a(v,r)\, b^2(v,r)\,dv^2 + 2 b(v,r)\, dv\, dr +
r^2 \left( d\theta^2 + \sin^2 \theta\, d\varphi^2 \right).
\]
While $b$ is required to be strictly positive,
$a$ can change sign, and as long as $b>0$
this metric is non-degenerate with signature $(-{} +{} +{} +{})$.
Asymptotic flatness means that the metric quantities $a$ and $b$
satisfy the boundary conditions
\begin{equation}\label{boundc}
   \lim_{r\to\infty}a(v, r)=\lim_{r\to\infty}b(v, r)=1.
\end{equation}
For a metric of this form the non-trivial components
of the Einstein equations
\[
G_{\alpha\beta} = 8 \pi T_{\alpha\beta}
\]
are found to be
\bea
-\frac{b}{r^2} \Bigl( r \partial_v a +
r a b \partial_r a + a^2 b - a b \Bigr)
&=&
8 \pi T_{00},\qquad\quad \label{E00}\\
\frac{b}{r^2} \Bigl(r \partial_r a + a -1\Bigr)
&=&
8 \pi T_{01},\label{E01}\\
\frac{2}{r b} \partial_r b
&=&
8 \pi T_{11},\label{E11}\\
\frac{r^2}{2b^2}
\Bigl(2\partial_{rv} b + b^2 \partial_{rr} a + 2 a b \partial_{rr} b
- \frac{2 \partial_r b \partial_v b}{b} + 3 b \partial_r a \partial_r b
&& \nonumber \\
+
\frac{2 b^2 \partial_r a}{r}+\frac{2 a b\partial_r b}{r}\Bigr)
&=&
8 \pi T_{22};
\label{E22}
\eea
the also non-trivial $33$ component is a multiple of the $22$ component.

In a collisionless gas the world lines of the particles are timelike geodesics.
The ensemble can be described by a non-negative number density
function $f$ on the tangent bundle $TM$
or equivalently on the cotangent bundle $TM^\ast$ of the spacetime.
The latter choice turns out to be advantageous for our analysis.
We denote by $(p^0,p^1,p^2,p^3)$ the canonical momenta corresponding
to the coordinates $(x^0,x^1,x^2,x^3)=(v,r,\theta,\varphi)$ so that
$(v,r,\theta,\varphi,p_0,p_1,p_2,p_3)$ coordinatize the cotangent bundle
$TM^\ast$.
On $TM^\ast$ the geodesic equations take the form
\[
\frac{dx^\alpha}{d\tau} = p^\alpha = g^{\alpha \beta}p_\beta,\
\frac{dp_\alpha}{d\tau} =
- \frac{1}{2} \frac{\partial g^{\beta\gamma}}{\partial x^\alpha}
p_\beta p_\gamma ,
\]
where $g^{\alpha\beta}$ is the inverse of the Lorentz
metric $g_{\alpha\beta}$.
Due to spherical symmetry the angular momentum
\[
L:= (p_2)^2 + \frac{1}{\sin^2\theta} (p_3)^2
\]
is conserved along geodesics,
\[
\frac{dL}{d\tau} = 0.
\]
We shall as usual assume that all the particles in the ensemble
have the same rest mass which we normalize to unity. Then their
density $f$ is supported on the mass shell defined by
\[
-1 = g^{\alpha\beta} p_\alpha p_\beta = \frac{2}{b} p_0 p_1 +
a\,(p_1)^2 +\frac{L}{r^2}.
\]
This implies that $p_1 \neq 0$ always, and since we want
$dv/d\tau = p^0 = p_1/b >0$,
i.e., all particles move forward in advanced time, we require $p_1 >0$
and can express $p_0$ as
\be \label{p_0}
p_0 = -\frac{b}{2}\left(a\, p_1 + \frac{1 + L/r^2}{p_1}\right).
\ee
Due to spherical symmetry the particle density $f$ is a function
of the variables $(v,r,p_1,L)$. Rewriting the relevant components of the
geodesic equations using $v$ as the parameter we find the characteristic
system (cf.\ (\ref{drdv}), (\ref{dp1dv}) below) of the
first order conservation law, i.e., the Vlasov equation, satisfied
by $f$:
\bea \label{vlasov}
&&
\partial_v f +
\frac{b}{2}\left( a - \frac{1 + L/r^2}{(p_1)^2}\right)\,
\partial_r f \nonumber \\
&&
\qquad\qquad
+ \frac{1}{2} \left(\frac{2 b L}{r^3 p_1} - \partial_r (a\,b)\, p_1 -
\partial_r b\, \frac{1 + L/r^2}{p_1}\right)
\partial_{p_1} f
=0.\quad
\eea
In order to close the system we have to define the energy momentum
tensor in terms of $f$ and the metric. In general,
\[
T_{\alpha\beta} = |g|^{-1/2}\int p_\alpha p_\beta f \frac{dp_0 dp_1 dp_2 dp_3}{m},
\]
where $|g|$ denotes the modulus of the determinant of the metric
and $m$ the rest mass of the particle with coordinates $(x^\alpha,p_\beta)$.
In the above coordinates and using the restriction to the mass shell,
\bea
T_{00}(v,r)
&=&
\frac{\pi}{r^2}
\int_0^\infty\int_0^\infty \frac{(p_0)^2}{p_1} f(v,r,p_1,L) \, dL\, dp_1,
\label{T00}\\
T_{01}(v,r)
&=&
\frac{\pi}{r^2}
\int_0^\infty\int_0^\infty p_0 f(v,r,p_1,L) \, dL\, dp_1,
\label{T01}\\
T_{11}(v,r)
&=&  \frac{\pi}{r^2}
\int_0^\infty\int_0^\infty p_1 f(v,r,p_1,L) \, dL\, dp_1,
\label{T11}\\
T_{22}(v,r)
&=&
\frac{\pi}{2 r^2}
\int_0^\infty\int_0^\infty \frac{L}{p_1} f(v,r,p_1,L) \, dL\, dp_1,
\label{T22}
\eea
where $p_0$ has to be expressed via (\ref{p_0}).

As our main result for the spherically symmetric and asymptotically flat
Einstein-Vlasov system (\ref{boundc})--(\ref{T22}) we prove that data given
at $v=0$ which satisfy certain explicit conditions and do not
contain a trapped surface, i.e., $a(0,\cdot) > 0$, launch solutions
where $a(v,r) < 0$ for some $v>0,ß r> 0$, i.e., a trapped surface forms.
A consequence of this result is that weak cosmic censorship holds for our
initial data in view of the results \cite{D05,DR05}.
The solutions we construct have the additional property that for $v$ sufficiently
large, all the matter is strictly within $\{r < 2 M\}$ and the generator
of the event horizon, which coincides with $r= 2 M$ for $v$ large,
is future complete.
The structure of the data at $v=0$ is essentially that
a static, $v$ independent state is surrounded by a shell of
matter, and the particles in this shell move towards
the center in a specified way.
We also study solutions which at $v=0$ contain a black hole of mass
$0 < m_0 < M$ where $M$ denotes the ADM mass for the complete spacetime.
The data are such that the initial black hole is irradiated by Vlasov
matter so that the apparent horizon is at $r= 2 m_0$ for
small advanced time $v$
and then grows from $r= 2 m_0$ to $r= 2 M$
in finite advanced time. The event horizon coincides with $r= 2 M$
for large $v$ but lies strictly between $r= 2 m_0$ and $r= 2 M$
for small $v$ so in particular, apparent horizon and event horizon
do not coincide. This behavior is analogous to what is found in
certain Vaidya spacetimes where matter, however, is described by
the ad-hoc model of null dust, cf.\ \cite[5.1.8]{P}.

Our main result described above should be related to a previous study
by Rendall \cite{Rend92} where he shows that there exist initial data
for the spherically symmetric Einstein-Vlasov system such that trapped
surfaces form in the evolution. However, the proof in \cite{Rend92} rests
on a continuity argument and it is not possible to tell whether a given
initial data set will evolve into a spacetime containing a trapped surface
which is in contrast to the explicit conditions derived in the present paper.

The paper proceeds as follows. In the next section we establish a
local existence and uniqueness result for the initial value
problem to the system (\ref{boundc})--(\ref{T22}), together with a
criterion which allows the solution to be extended as long as certain
quantities are controlled. In addition we collect some general
information on the solutions to be used in the further analysis.
In Section~\ref{secpbounds} we show that for particles
in the outer shell $p_1$ is bounded
both from above and away from zero as long as the shell stays away
from the center. This result is used in Section~\ref{sectrapped}
to prove the formation of a trapped surface out of regular data.
An essential step here is to show that if the particles in the outer
shell move towards the center initially in a specified way they continue
to do so for $v>0$. In the last section we study the solutions after
a trapped surface has formed and also obtain the Vaidya type spacetimes
mentioned above.

We conclude this introduction with some further references
to the literature. Background on the Einstein-Vlasov system
and relativistic kinetic theory can be found in \cite{And05, Rend97}.
A distinguishing feature of the Vlasov matter model is that
in the Newtonian case, i.e., for the Vlasov-Poisson system,
no gravitational collapse occurs and solutions exist globally
in time, cf.\ \cite{LP,Pf}. For more background on the
Vlasov-Poisson system we refer to \cite{Rein07}.
\section{Local existence, conservation laws, and a-priori-bounds}
\setcounter{equation}{0}
In this section we collect a number of observations, estimates, and
a-priori-bounds which will play a role in what follows and which
also lead to a local existence result and a continuation criterion.

To begin with, let us assume that a non-negative, compactly supported
number density $f(v)  \in C^1_c(]0,\infty[^2\times [0,\infty[)$ is given at some
instant $v\geq0$.
We want to show that at that instant the metric is then determined
explicitly in terms of $f(v)$. Firstly, the field equation (\ref{E11}) can be
integrated to yield
\be \label{bformula}
b(v,r) = \exp\left(-4\pi \int_r^\infty \eta T_{11}(v,\eta)\, d\eta \right).
\ee
Since the formula (\ref{T11}) for $T_{11}$ does not contain a
metric coefficient, (\ref{bformula}) defines $b(v)$ in terms of
$f(v)$, and
\be \label{best}
0 < b(v,r) \leq 1,\ r > 0,
\ee
in particular $b$ is positive as required. The
boundary condition (\ref{boundc}) holds as well, indeed
$b(v,r) = 1$ when $r$ lies to the right of $\supp f(v)$.
In order to express $a$ we observe that
\be \label{T01S}
T_{01}
=
-\frac{b}{2} \left(a \, T_{11} + S\right)
\ee
where
\be \label{S}
S :=
\frac{\pi}{r^2}
\int_0^\infty\int_0^\infty \frac{1+L/r^2}{p_1} f \, dL\, dp_1 .
\ee
The field equation (\ref{E01}) can now be rewritten
in the form
\be \label{alpharrewr}
\partial_r (r a-r) + 4 \pi r T_{11}\; (ra-r) = - 4 \pi r^2 (T_{11}+S).
\ee
Since $(ra-r)_{|r=0} = 0$ this can be integrated to give
\be \label{alphaformula}
a(v,r) =
1- \frac{1}{r} \int_0^r 4\pi \eta^2 (T_{11}+S)
\exp\left(-\int_\eta^r 4 \pi \sigma T_{11}
d\sigma\right)\, d\eta;
\ee
clearly $a (v,\infty)=1$.
Hence given $f(v)$ at some instant $v$ the metric is explicitly
determined. Notice that when in Section~\ref{afterts} we consider
the system with a black hole at the center we replace
the above boundary condition for $r a - r$ at $r=0$ by one
at $r=\infty$.

If the metric coefficients $a$ and $b$ are
given and sufficiently smooth on some interval $[0,V[$
we denote by $(R,P_1)(s,v,r,p_1,L)$ the solution of the
characteristic system
\bea
\frac{dr}{ds}
&=&
\frac{b}{2}\left( a - \frac{1 + L/r^2}{(p_1)^2}\right),\label{drdv}\\
\frac{dp_1}{ds}
&=&
\frac{1}{2} \left(\frac{2 b L}{r^3 p_1} - \partial_r (a\,b)\, p_1 -
\partial_r b\, \frac{1 + L/r^2}{p_1}\right) \label{dp1dv}
\eea
of the Vlasov equation (\ref{vlasov})
with $(R,P_1)(v,v,r,p_1,L)= (r,p_1)$.
Then
\be \label{fformula}
f(v,r,p_1,L) = \fn \left((R,P_1)(0,v,r,p_1,L),L\right)
\ee
is the solution of the Vlasov equation satisfying
the initial condition $f_{|v=0} = \fn$.
If $\fn$ is non-negative and compactly supported
then these properties are inherited by $f(v)$.

The above observations allow for the following iterative scheme.
If initial data $\fn$ are given we define
$f_0 (v,r,p_1,L) := \fn (r,p_1,L)$.
If $f_n$ is given we define $T_{11,n}$ and $S_n$ by substituting
$f_n$ into the formulas (\ref{T11}) and (\ref{S}). Next we define
$a_n$ and $b_n$ through
(\ref{bformula}) and (\ref{alphaformula}).
Finally, we obtain the next iterate $f_{n+1}$ via (\ref{fformula}),
using $a_n$ and $b_n$ in (\ref{drdv}), (\ref{dp1dv}).
This iterative scheme leads to the following local existence result.
\begin{theorem} \label{locex}
Let $\fn \in C^1_c(]0,\infty[^2 \times [0,\infty[)$
be compactly supported and non-negative.
Then there exists a unique solution
$f \in C^1([0,V[ \times ]0,\infty[^2 \times [0,\infty[)$ of the system
(\ref{boundc})--(\ref{T22})
with $f_{\mid v=0} = \fn$ where $V>0$. Let $V$ be chosen maximal.
If
\[
\sup\left\{p_1 + \frac{1}{p_1} + \frac{1}{r}  \mid
(r,p_1,L) \in \supp f(v),\ 0 \leq v < V \right\} < \infty
\]
then $V=\infty$.
\end{theorem}
\prf
In Schwarzschild coordinates the analogous proof has been carried
out in detail in \cite{Rein95,RR1}. Here we only address some
key issues. Firstly, by construction the iterative scheme indicated
above will converge to a solution of the subsystem where
only the field equations (\ref{E01}) and (\ref{E11}) hold.
The remaining field equations can then be derived exploiting the
following observation. If $\nabla_\alpha$ denotes the covariant derivative
corresponding to the given metric, then the Vlasov equation
implies that $\nabla_\alpha T^{\alpha \beta} = 0$. Also the Einstein
tensor satisfies the relation $\nabla_\alpha G^{\alpha \beta} = 0$.
The relation $\nabla_\alpha E^{\alpha \beta} = 0$ satisfied by
$E^{\alpha \beta} := G^{\alpha \beta} - 8 \pi T^{\alpha \beta}$
then implies that the remaining field equations hold.

Clearly, the supremum $Q$ in the statement of the theorem
together with (\ref{fformula}) controls $T_{11}$ and $S$,
hence $a$ and $b$, and via the corresponding field
equations also their first order derivatives
$\partial_r a, \partial_r b, \partial_v a$; notice that
\be \label{alphav}
-\frac{b}{r} \partial_v a = G_{00} + a\, b\, G_{01}
= 8 \pi (T_{00} + a\, b\, T_{01}).
\ee
But in order to extend $f$ as a $C^1$ solution
it is necessary to control the derivatives of $R$ and $P_1$ with
respect to their data $r,p_1,L$ which seems to require second
order derivatives of $a$ and $b$ which are not controlled by the
supremum $Q$. However, the quantities
\[
\xi = \partial_r R,\
\eta = \partial_r P_1 - P_1 \frac{\partial_r b}{b} \partial_r R
\]
satisfy a system of ordinary differential equations
where second order derivatives of the metric coefficients
appear only in the combination
\[
2\partial_{rv} b + b^2 \partial_{rr} a + 2 a b \partial_{rr} b
- \frac{2 \partial_r b \partial_v b}{b}
\]
which appears in the field equation (\ref{E22}) and is therefore
controlled by $Q$ as well; notice that $\partial_v b$ which appears
in this combination cannot be controlled by itself in terms of $Q$.
A similar argument also helps to obtain the bounds on the iterates
required for their convergence.
The differential geometric background
of this maneuver is that the evolution of the
derivatives of characteristics
with respect to their data is governed by the geodesic deviation
equation where derivatives of the metric enter only through
the Riemann curvature tensor, and due to the symmetry the
latter contains second order derivatives of the metric coefficients
in the same combination as they appear in the Einstein tensor and
hence in the field equations.
\prfe

It seems necessary that the support of the matter for the solution
and hence for the initial data is bounded away from the origin.
At first glance this is due to using polar coordinates
$(r,\theta,\varphi)$, and one may hope to cure this by passing to
the induced Cartesian coordinates
$r (\sin\theta \cos\varphi,\sin\theta \sin\varphi,\cos\theta)$,
as was done in \cite{Rein95,RR1}.
However, in the case of Eddington-Finkelstein
coordinates this does not seem to help;
the metric, when written in Cartesian
coordinates, is not $C^2$ at the origin, indeed, not even continuous.
Since this issue does not play a role in out analysis we do not pursue
it further.

In what follows below, two situations will be considered.
In the first we place a static, $v$-independent
solution in the center and
surround it by a shell of non-static Vlasov matter. Since
we will only need to analyze this case as long as the outer
non-static matter does not reach the central static part,
the considerations above are sufficient to deal with it.
In the second case, studied in Section~\ref{afterts},
we consider the system only for $r\geq r_0$ where $a(0,r_0) < 0$,
so again a neighborhood of the center is avoided.

Coming back to the first case we need to see that the system
indeed has static, $v$-independent solutions. The existence
of static, spherically symmetric and compactly supported
solutions has been established in \cite{Rein94,RR3,RR4}
using Schwarzschild coordinates. Given such a static solution
in Schwarzschild coordinates its metric takes the form
\[
ds^2=-e^{2\mu(r)}dt^2+e^{2\lambda(r)}dr^2+
r^2(d\theta^2+\sin^2\theta\,d\varphi^2).
\]
Changing to the variable
$v=v(t,r) = t + \int_0^r e^{\lambda(\eta) - \mu(\eta)}d\eta$
brings this metric into the Eddington-Finkelstein form
where
\[
a = e^{-2\lambda},\ b = e^{\lambda + \mu}
\]
so that these transformed metric coefficients
are independent of $v$ as desired, and the same is then true for
the particle density $f$ of the steady state.

In what follows $f_s$ denotes a fixed spherically symmetric steady state
of the Einstein-Vlasov system with spatial support
in $[0,r_0]$. We consider initial data of the form
\be \label{datasc}
\fn := f_s +\fn_\mathrm{out}
\ \mbox{where}\
\fn_\mathrm{out} \in C^1_c(]r_0,\infty[ \times ]0,\infty[ \times [0,\infty[),\
\fn_\mathrm{out}\geq 0.
\ee
For a solution launched by such data we introduce the notation
\be \label{foutdef}
f_\mathrm{out}(v,r,p_1,L):= \fn_\mathrm{out}((R,P_1)(0,v,r,p_1,L),L),
\ee
i.e., $f_\mathrm{out}$ is the density of the particles which initially
do not belong to the steady state. Due to spherical symmetry there
are no gravitational waves, and the only way that $f_\mathrm{out}$
can influence the matter in the region $\{ r < r_0\}$ is if
outer particles actually reach this region. As long as this does not
happen the matter
in the central region stays in its equilibrium configuration.
Hence we obtain the following
corollary to Theorem~\ref{locex}.
\begin{cor} \label{locexsc}
Initial data $\fn$ as specified in (\ref{datasc}) launch
a unique solution
$f\in C^1([0,V[ \times ]0,\infty[^2 \times [0,\infty[)$ of the system
(\ref{boundc})--(\ref{T22})
with $f_{\mid v=0} = \fn$ where $V>0$. Let $V$ be chosen maximal.
If
\[
\sup\left\{p_1 + \frac{1}{p_1} \mid
(r,p_1,L) \in \supp f_\mathrm{out} (v),\ 0 \leq v < V \right\} < \infty
\]
and
\[
\inf\left\{r \mid (r,p_1,L) \in \supp f_\mathrm{out} (v),\
0 \leq v < V \right\} > r_0
\]
then $V=\infty$.
\end{cor}
We now collect some properties of the local solutions
which are used in what follows. First we look at the
ADM mass and resulting a-priori bounds.
\begin{proposition} \label{conslaws}
Let
\[
m(v,r) := \frac{r}{2} \left(1-a(v,r)\right),\ \mbox{i.e.,}\
a(v,r) = 1 - \frac{2 m(v,r)}{r}.
\]
Then the quasi-local ADM mass $m$ is given by
\bea \label{mform1}
m(v,r)
&=&
2 \pi \int_0^r \eta^2 (T_{11}+S)
\exp\left(-\int_\eta^r 4 \pi \sigma T_{11}
d\sigma\right)\, d\eta \nonumber\\
&=&
2 \pi \int_0^r \eta^2 (T_{11}+S)(v,\eta)
\frac{b(v,\eta)}{b(v,r)}\, d\eta.
\eea
For $r$ sufficiently large,
\[
m(v,r) = M = \lim_{r\to \infty} m(v,r)
\]
which is a conserved quantity, the ADM mass. Moreover,
\be \label{mform2}
m(v,r)
=
2 \pi \int_0^r \eta^2 (a T_{11}+S)(v,\eta)\, d\eta,
\ee
and
\bea
a(v,r)
&\leq&
1,\ \nonumber \\
b(v,r)\ m(v,r)
&=&
2 \pi \int_0^r \eta^2 (T_{11}+S) b\, d\eta \nonumber \\
&\leq&
2 \pi \int_0^\infty \eta^2 (T_{11}+S) b\, d\eta = M. \label{mapri}
\eea
\end{proposition}
\prf
The formula (\ref{mform1}) for $m(v,r)$
results from comparing its relation to $a$
with (\ref{alphaformula}) and (\ref{bformula}).
Recalling (\ref{alphav}),
\be \label{mv}
\partial_v m
=
-\frac{r}{2} \partial_v a =
4 \pi r^2 \left(\frac{1}{b}T_{00} + a T_{01}\right)
\ee
in particular, $m(\cdot,r)$ is constant for $r$ large
enough and hence to the right of the support of the matter.
If for the moment we denote the right hand
side of (\ref{mform2}) by $\tilde m$ then clearly
$\tilde m(v,0) = 0 = m(v,0)$ , and the
differential equation (\ref{alpharrewr}) for $r a - r$ is equivalent to
the fact that $\partial_r \tilde m = \partial_r m$.
By (\ref{mform1}), $m\geq 0$ and hence $a \leq 1$,
and (\ref{mform1}) together with the conservation of $M$
and the fact that $b(v,\infty)=1$ implies (\ref{mapri}).
\prfe
To conclude this section we collect some information on
sign changes in $a$.
\begin{proposition} \label{aneg}
If $a(v_0,r_0) < 0$ for some $v_0\geq 0, r_0>0$ then  $a(v,r_0) < 0$
for all $v>0$ for which the solution exists. If $a(v,r) < 0$
then all timelike or null geodesics at the spacetime point
$(v,r,\theta,\phi)$ move towards strictly smaller values of $r$.
\end{proposition}
\prf
First we note that by (\ref{mv}),
\beas
\partial_v a
&=&
- 8 \pi r \left(\frac{1}{b} T_{00} +a T_{01}\right)\\
&=&
- \frac{2 \pi^2 b}{r} \int_0^\infty \int_0^\infty \frac{1}{(p_1)^3}
\left(\left(1+L/r^2\right)^2 - a^2 (p_1)^4\right) f \,dL\,dp_1.
\eeas
Hence $\partial_v a(v,r)=0$ if $f(v,r,\cdot) = 0$.
If $f(v,r,\cdot) > 0$ then $\partial_v a(v,r)<0$,
provided $a(v,r)$
is sufficiently close to $0$ so that the term in parentheses is positive;
notice that $f(v,r, \cdot)$ has compact support. Hence if $a(v_0,r_0) < 0$
at some $v_0,r_0$ this sign must be preserved for all $v> v_0$.
The remaining assertion follows from the geodesic equations.
\prfe

\section{A lower and an upper bound on $p_1$}
\label{secpbounds}
\setcounter{equation}{0}
In this section we consider solutions launched by
data as specified in (\ref{datasc}), i.e., with a steady
state at the center.
The aim is to prove that
$p_1$ is bounded from above and below on the support
of $f_\mathrm{out}$ where we recall (\ref{foutdef}).
In connection with the continuation criterion
in Corollary~\ref{locexsc} this means that a solution can
only blow up if particles from the exterior mass shell
reach the central region $\{r<r_0\}$ where the steady
state part is supported. The result may be compared to
\cite{RRS} where it is shown that in Schwarzschild coordinates
solutions which blow up at all must do so at the center first.
\begin{theorem} \label{pbounds}
Let $V>0$ be such that $r>r_0$ for $(r,p_1,L) \in \supp f_\mathrm{out}(v)$
and $0\leq v \leq V$.
Then there are constants $C_1, C_2 > 0$ such that
$C_1\leq p_1\leq C_2$ for $(r,p_1,L) \in \supp f(v)$
and $0\leq v \leq V$.
\end{theorem}
In the proof of this theorem we need the following a-priori-bound.
\begin{lemma}\label{apri2}
Let $V>0$ and $r_0 > 0$ be such that $f(v,r_0,\cdot,\cdot)=0$
for $0\leq v \leq V$.
Then
\[
\int_{r_0}^{\infty} r^2 T_{11}(v,r)\,dr
\leq
\int_{r_0}^{\infty} r^2 T_{11}(0,r)\,dr
+ \frac{M}{2 \pi r_0} v,\ 0\leq v \leq V.
\]
\end{lemma}
\prf
Using the Vlasov equation, integration by parts, and the field equations
we find that
\[
\frac{d}{dv}\int_{r_0}^{\infty} r^2 T_{11}\,dr
=
-\frac{1}{2}\int_{r_0}^{\infty}br T_{11}(1-a)dr
+ \pi \int_{r_0}^{\infty}\int_0^{\infty}\int_0^{\infty}
\frac{bL}{r^3 p_1} f \, dL\, dp_1\,dr.
\]
Now $a\leq 1$ and $T_{11} \geq 0$ so the first term
can be dropped. As to the second, we observe that
\[
\frac{bL}{r^3 p_1} \leq b \frac{1+L/r^2}{p_1} \frac{1}{r},
\]
and recalling the expression (\ref{S}) for $S$  we obtain the estimate
\[
\frac{d}{dv}\int_{r_0}^{\infty}r^2 T_{11}\,dr
\leq
\int_{r_0}^{\infty} r b S dr
\leq
\frac{1}{r_0} \int_{r_0}^{\infty} r^2 b S dr.
\]
By (\ref{mapri}),
\[
M =  2 \pi \int_0^\infty r^2 (T_{11} + S) \,b(v,r)\, dr
\geq
2 \pi  \int_{r_0}^\infty r^2 S(v,r)\, b(v,r)\, dr,
\]
hence
\[
\frac{d}{dv}\int_{r_0}^{\infty}r^2 T_{11}\,dr \leq \frac{M}{2 \pi r_0}
\]
and the claim of the lemma follows.
\prfe

\noindent
{\bf Proof of Theorem~\ref{pbounds}.}
Let $[0,V] \ni s \mapsto (r(s),p_1(s),L)$ denote a characteristic
in $\supp f_\mathrm{out}$. Using (\ref{drdv}) we can rewrite (\ref{dp1dv})
in the form
\begin{equation}\label{p1evol}
\frac{d}{ds}p_1=\frac{bL}{r^3p_1}-\left(\frac{1}{2}\partial_r a b
+ a \partial_r b -\frac{\partial_r b}{b} \frac{dr}{ds}\right)\, p_1.
\end{equation}
In order to obtain a lower bound for $p_1$ we observe that
\beas
\frac{d}{ds}\frac{1}{p_1}
&=&
- \frac{bL}{r^3(p_1)^3}
+\left(\frac{1}{2}\partial_r a b
+ a \partial_r b -\frac{\partial_r b}{b} \frac{dr}{ds}\right)
\, \frac{1}{p_1}\\
&\leq&
\left(\frac{1}{2}\partial_r a b
+ a \partial_r b -\frac{\partial_r b}{b} \frac{dr}{ds}\right)
\, \frac{1}{p_1}.
\eeas
Consider an arbitrary instant of advanced time $0<v_0\leq V$.
Applying Gronwall's lemma we find that we need to estimate the integral
\[
\int_0^{v_0} \left[\left(\frac{1}{2}\partial_r a b
+ a \partial_r b\right)(s,r(s)) -\left(\frac{\partial_r b}{b}\right)(s,r(s))
\frac{dr}{ds}(s)\right]\, ds
\]
which is a curve integral along the curve
\[
\gamma=\{(s,r(s))\mid 0\leq s\leq v_0\}.
\]
In order to estimate this integral we apply Green's formula in the plane.
For this we define
\beas
C_1
&=&
\{(v_0,r) \mid r(v_0)\leq r\leq R\},\\
C_2
&=&
\{(v,R) \mid 0\leq v\leq v_0\},\\
C_3
&=&
\{(0,r) \mid r(0)\leq r\leq R\}.
\eeas
Here $R>0$ is large, and we will let $R\to\infty$.
We orient the closed curve $\Gamma=\gamma+C_1+C_2+C_3$ clockwise.
Now
\beas
\int_0^{v_0} \left[\frac{1}{2}\partial_r a b
+ a \partial_r b - \frac{\partial_r b}{b}
\frac{dr}{ds}\right]\,ds
&=&
\int_\gamma \left[\left(\frac{1}{2}\partial_r a b
+ a \partial_r b\right)\,dv -\left(\frac{\partial_r b}{b}\right)\,dr \right]
\nonumber \\
&=&
\oint_\Gamma -\int_{C_1} -\int_{C_2} -\int_{C_3}.
\eeas
We denote by $\Omega$ the domain enclosed by $\Gamma$ and apply
Green's formula in the plane to find that
\beas
\oint_\Gamma \left[ \cdots \right]
&=&
-\frac{1}{2} \iint_{\Omega}
\left[\frac{\partial}{\partial v}\left(\frac{2 \partial_r b}{b}\right) +
\frac{\partial}{\partial r}
\left(\partial_r a b + 2 a \partial_r b \right)\right]\,dr\,dv\\
&=&
- \iint_{\Omega}\left[\frac{b}{r^2} G_{22} -
\frac{a\,\partial_r b}{r}-\frac{\partial_r a\, b}{r}\right]\,dr\,dv;
\eeas
recall the field equation (\ref{E22}) for the $22$-component $G_{22}$
of the Einstein tensor. If we now use the field equations and the definitions
of the energy momentum tensor it turns out that
\[
\frac{b}{r^2} G_{22} -
\frac{a\,\partial_r b}{r}-\frac{\partial_r a\, b}{r}
=
\frac{4 \pi^2}{r^2}b \int_0^\infty \int_0^\infty
\frac{1+2 L/r^2}{p_1} f \, dL\, dp_1
- \frac{b}{r^2} \frac{2 m}{r}.
\]
Hence by (\ref{mapri}),
\[
\oint_\Gamma\cdots \leq  \iint_{\Omega} \frac{2 m b }{r^3}\, dr\, du
\leq \frac{M}{r_0^2}v_0.
\]
Next we use the field equation (\ref{E11}) and Lemma~\ref{apri2}
to get the estimate
\bea \label{-C1est}
-\int_{C_1} \cdots
&=&
\int_{r(v_0)}^R \frac{\partial_r b(v_0,r)}{b(v_0,r)} dr
\leq
4 \pi \int_{r_0}^\infty r T_{11}(v_0,r)\, dr \nonumber \\
&\leq&
\frac{1}{r_0} 4 \pi \int_{r_0}^\infty r^2 T_{11}(v_0,r)\, dr \\
&\leq&
\frac{1}{r_0}\left(4 \pi \int_{r_0}^\infty r^2 T_{11}(0,r)\, dr
+ \frac{2 M}{r_0} v_0\right) . \nonumber
\eea
The $C_2$-contribution vanishes in the limit $R \to \infty$:
\[
-\int_{C_2} \cdots
=
\int_0^{v_0}\left(\frac{1}{2}\partial_r a b
+ a \partial_r b\right)(v,R)\,dv \leq \frac{M}{R^2} v_0,
\]
since for $r$ sufficiently large and in particular outside the support
of $f$, $b= 1$ and $\partial_r a = 2 M/r^2$. Finally,
\[
-\int_{C_3} \cdots = - \int_{r(0)}^R \frac{\partial_r b(0,r)}{b(0,r)} dr  \leq 0.
\]
Altogether this implies, using Gronwall's lemma, the estimate
\be \label{1/p1est}
\frac{1}{p_1(v)}
\leq
\frac{1}{p_1(0)}
\exp\left(\frac{4 \pi}{r_0} \int_{r_0}^\infty r^2 T_{11}(0,r)\, dr
+ \frac{3 M}{r_0^2} v\right).
\ee
In order to estimate $p_1$ from above we start from (\ref{p1evol}). The term
$(bL)/(r^3 p_1)$ is bounded by the previous estimates so after using
Gronwall's lemma we have to estimate the term
\[
- \int_\gamma \left[\left(\frac{1}{2}\partial_r a b
+ a \partial_r b\right)\,dv -\left(\frac{\partial_r b}{b}\right)\,dr \right]
=
- \oint_\Gamma + \int_{C_1} + \int_{C_2} + \int_{C_3}
\]
from above, where the curves are defined and oriented as before.
The $C_1$-contribution is now negative and can be dropped. Since for
$r$ sufficiently large and outside the support of $f$, $b=1$
and $\partial_r a \geq 0$ the $C_2$-contribution can be dropped as well.
The $C_3$-contribution is determined by the initial data, so it remains
to estimate the integral over the closed curve $\Gamma$ which we again
turn into an integral over $\Omega$ using Green's formula. Due to the
change in sign we are left with estimating the integral
\[
\iint_\Omega \frac{4 \pi^2}{r^2}b \int_0^\infty \int_0^\infty
\frac{1+2L/r^2}{p_1} f \, dL\, dp_1\,dr\,dv.
\]
Using the already established bounds this amounts to estimating the
integral $\int_{r_0}^\infty \int_0^\infty \int_0^\infty f \, dL\, dp_1 \,dr$
which represents the number of particles in the domain $\{ r \geq r_0\}$
and is conserved for $v\in [0,V]$.
The proof of Theorem~\ref{pbounds} is complete.
\prfe
Together with the continuation criterion from the local existence result
Corollary~\ref{locexsc} we obtain the following corollary.
\begin{cor} \label{Vcor}
There exists $V>0$ such that the solution exists on the interval
$[0,V[$ and $r> r_0$ for all particles in the outer matter.
If $V$ is chosen maximal, then $V=\infty$ if
\[
\inf\{ r \mid (r,p_1,L) \in \supp f_\mathrm{out}(v),\ 0 \leq v < V\} >r_0,
\]
i.e., the solution can be extended as long as the outer matter stays outside
$\{ r \leq r_0\}$.
\end{cor}
\section{The formation of a trapped surface}
\label{sectrapped}
\setcounter{equation}{0}
In this section we want to specify conditions on the data
such that a trapped surface evolves, but is not already
present in the data. The data are again of the form (\ref{datasc}),
i.e., they have a steady state $f_s$ at the center
whose mass we denote by $m_0$. We fix some notation:
\be \label{suppfn}
\supp \fn_\mathrm{out} \subset [R_0,R_1] \times [p_- ,p_+] \times [0,L_+]
\ee
where
\[
0 < r_0 < R_0 < R_1,\ 0 < p_- < p_+,\ L_+ >0,
\]
and $M > m_0$ is the mass of $\fn := f_s + \fn_\mathrm{out}$.
If $R_0 \geq 2 M$ then the data do not contain a trapped surface
in the sense that $a(0,\cdot) >0$.
We define
\be \label{Pdef}
P := - \max \left\{p^1 \mid (r,p_1,L) \in \supp \fn_\mathrm{out}\right\}.
\ee
Since
\be \label{p^1eq}
p^1
=
\frac{1}{b} p_0 + a p_1 =
\frac{1}{2} a p_1 - \frac{1}{2} \frac{1+L/r^2}{p_1}
\leq
\frac{1}{2} p_1 - \frac{1}{2} \frac{1}{p_1}
\ee
we have $P > 0$, i.e., all particles move inward initially,
if for example $p_+ < 1$.
The following theorem is the main result of the present paper.

\begin{theorem} \label{ts}
Let data $f_s + \fn_\mathrm{out}$ be given such that
\be \label{suppcond}
R_0 \geq 2 M >r_0,\ L_+ := 12 m_0^2,\
2 P \geq 1 + \frac{L_+}{r_0^2},
\ee
and such that there exists $V > 0$ with the property that
\bea
\frac{2 P^2}{1+L_+/r_0^2} \exp\left(-\frac{2 M}{r_0}\right)\,  V
&>&
R_1 - 2 M, \label{tsform}\\
\frac{1}{2}\left(1+\frac{L_+}{r_0^2}\right) \frac{1}{p_-^2}
\exp\left(\frac{2 M}{r_0^2} V + \frac{4 M}{r_0}\right) \, V
&<&
R_0 - r_0. \label{nointerference}
\eea
Then the solution launched by $f_s + \fn\,$ forms a trapped surface
at some advanced time $v< V$.
\end{theorem}
{\bf Remark.}
Note that this result implies that weak cosmic censorship holds for
these data in view of the results \cite{D05,DR05}.
In Section~\ref{afterts} we will see that for data as considered above
an event horizon evolves which is future complete,
cf.\ Theorem~\ref{asympt} and Corollary~\ref{corasympt}.

\smallskip

\noindent
{\bf Remark.} It is easy to see that data which satisfy
the conditions above do exist: First we arbitrarily fix
the central steady state $f_s$ and hence also $r_0$ and $m_0$.
Next we fix $R_0, M, p_+$ in such a way that (\ref{suppcond})
holds; notice that $P$ becomes large if we chose $p_+>0$ small,
cf.\ (\ref{p^1eq}). Then we choose $0 < p_- < p_+$.
If we replace the inequality in (\ref{nointerference}) by an
equality this uniquely determines $V$ which we then choose slightly
smaller to preserve the inequality. Substituting this $V$ into
(\ref{tsform}) we see that that relation is satisfied provided
$R_1$ is sufficiently close to $2 M$,
and all the support parameters
and the mass of $\fn_\mathrm{out}$ are fixed.
Now we fix any non-negative, non-vanishing function $g$
which satisfies the support
conditions, consider data $A g$ with some amplitude $A\geq 0$
and denote the induced ADM mass and metric coefficient
by $M_A$ and $a_A$ respectively. From
(\ref{mform1}) we see that $M_A$ depends continuously on
$A$ with $M_0=m_0$.
If $a_A$ remains positive as $A$ increases then (\ref{mform2})
shows that $M_A$ becomes as large as we wish with increasing $A$.
So assume that $a_{A^\ast}(r^\ast)=0$ for some $r^\ast > R_0$ and some
value of amplitude $A^\ast$, and $a_A$ is positive for all $A<A^\ast$.
Then $2 m_{A^\ast}(r^\ast) = r^\ast > R_0 \geq 2 M$ and $a_{A^\ast}$
is still non-negative, hence $M_{A^\ast}>M$, and $M_{A}=M$
for some smaller value of $A$.
Hence  $\fn_\mathrm{out}$ satisfying all support conditions
and having the proper mass exist.
Since $R_0 \geq 2 M$, $a(0,\cdot)>0$ for any such $\fn_\mathrm{out}$.
\smallskip

A major step in the proof of Theorem~\ref{ts}
is to show that, for suitable data,
the particles move inward as long as they stay outside $r=r_0$.
Since we will recycle this result in Section~\ref{afterts},
we formulate it in a more general setting than needed here.
\begin{lemma} \label{ingoing}
Let $V$ be the length of the maximal existence interval of a
solution $f$ as in Corollary~\ref{Vcor}, and assume that
\[
L_+ \leq 12 m_0^2\ \mbox{and}\ P > 0.
\]
Then
\[
\max\left\{p^1 \mid
(r,p_1,L) \in \supp f_\mathrm{out}(v),\ 0\leq v < V\right\}\leq
- P < 0.
\]
\end{lemma}
\prf
By continuity and the assumption on the data there is some
interval $[0,V^\ast[ \subset [0,V[$ such that
$p^1 < 0$ for all $(r,p_1,L) \in \supp f_\mathrm{out}(v)$
with $0\leq v < V^\ast$. We choose $V^\ast$ maximal and
have to show that $V^\ast = V$. To this end we consider
a characteristic in $\supp f_\mathrm{out}$ and
parametrize it by proper time.
After some computation we find that
\
\beas
\dot p^1
&=&
\frac{d}{d\tau} \left( \frac{1}{b}p_0 + a p_1\right)\\
&=&
- \frac{\partial_r b}{b^3} (p_0)^2 + \frac{\partial_v a}{2 b} (p_1)^2 +
\frac{a \partial_r a}{2} (p_1)^2 +
\frac{\partial_r a}{b} p_0 p_1 + a \frac{L}{r^3}\\
&=&
-\frac{4 \pi r}{b^2} T_{11} (p_0)^2
+ \left( \frac{8 \pi r}{b^2} T_{01} + \frac{1}{r b}(1-a)\right)\, p_0 p_1\\
&&
+\left(-\frac{4 \pi r}{b^2} T_{00} + \frac{a}{2r}(1-a)\right)\, (p_1)^2
+ a \frac{L}{r^3}\\
&=&
\frac{1}{r b} (1-a) p_0 p_1 + \frac{a}{2 r}(1-a) (p_1)^2
+ a \frac{L}{r^3} \\
&&
- \frac{4 \pi^2}{r b^2} \int_0^\infty \int_0^\infty
\left[\tilde p_1 (p_0)^2 - 2 \tilde p_0 p_0 p_1 +
\frac{(\tilde p_0)^2}{\tilde p_1} (p_1)^2\right]
\,  f\, d\tilde L\, d\tilde p_1,
\eeas
where $\tilde p_0$ is defined as in (\ref{p_0}) but in terms of
the integration variables
$\tilde p_1$ and $\tilde L$. Now
\[
[\ldots] =
\left[ \sqrt{\tilde p_1} p_0 -
\frac{\tilde p_0 p_1}{\sqrt{\tilde p_1}}\right]^2  \geq 0,
\]
so after substituting the definition (\ref{p_0}) for $p_0$,
\be \label{dotp^1}
\dot p^1
\leq
- \frac{1}{2 r}(1-a)\left(1+\frac{L}{r^2}\right) -(1-a) \frac{L}{r^3} +
\frac{L}{r^3}.
\ee
If $a(v,r) < 0$, then
\[
\dot p^1
\leq
- \frac{1}{2 r}\left(1+\frac{L}{r^2}\right) - \frac{L}{r^3}+ \frac{L}{r^3} < 0.
\]
If $a(v,r) \geq 0$, then we observe the relation
$a(v,r) = 1-2m(v,r)/r$ and investigate the behavior of
$m(v,r)$ with respect to $v$. By Proposition~\ref{aneg},
$a(\cdot,r) \geq 0$ on $[0,v]$.
By (\ref{mv}) and (\ref{T00}), (\ref{T01}),
\[
\partial_v m =
4 \pi^2\int_0^\infty\int_0^\infty
\left(\frac{1}{b}\frac{(p_0)^2}{p_1} + a p_0\right)\, f\, dL\, dp_1,
\]
and
\[
\frac{1}{b}\frac{(p_0)^2}{p_1} + a p_0
= \frac{p_0}{p_1} p^1.
\]
By assumption, $p^1 < 0$
on $[0,v]\subset [0,V^\ast[$ for all particles,
also $p_1 > 0$ for all particles, and
\[
p_0 = -\frac{b}{2}\left( a p_1 + \frac{1+L/r^2}{p_1}\right) < 0
\]
because of the sign of $a(\cdot,r)$ on $[0,v]$. Hence
$\partial_v m(\cdot, r)\geq 0$ on $[0,v]$, and
$m(v,r) \geq m(0,r) \geq m_0$.
Hence (\ref{dotp^1}) implies that
\beas
\dot p^1
&\leq&
-\frac{1}{2 r} \frac{2 m}{r} \left(1+\frac{L}{r^2}\right)
-\frac{2 m}{r}\frac{L}{r^3} + \frac{L}{r^3}\\
&\leq&
\frac{1}{r^4} \left(L r - 3 L m_0 - r^2 m_0 \right)
= \frac{1}{r^4}
\left( \frac{L^2}{4 m_0} -\left(\sqrt{m_0} r - \frac{L}{2 \sqrt{m_0}}\right)^2
- 3 L m_0\right)\\
&\leq&
\frac{L}{4 r^4 m_0} \left(L - 12 m_0^2\right) \leq 0.
\eeas
Together with the case $a(v,r) < 0$ this shows that
$\dot p^1\leq 0$ for any characteristic in $\supp f_\mathrm{out}(v)$
with $v\in ]0,V^\ast[$, hence $V^\ast = V$, and the proof is complete.
\prfe

\noindent
{\bf Proof of Theorem~\ref{ts}.}
Assume that $a (v,r) \geq 0$ for all $r > 0$ and $0 < v \leq V$ and
as long as the solution exists. We will show that the solution must then
exist on the interval $[0,V]$ and $m(V,r) = M$ for some $r<2 M$ so that
$a (V,r) < 0$ which is the desired contradiction. The basic idea
is that (\ref{nointerference}) guarantees that the continuation
criterion from Corollary~\ref{Vcor} applies to the interval $[0,V]$ so that
the solution exists there, while (\ref{tsform}) guarantees that
all the matter arrives inside $\{r < 2 M\}$ within that interval
of advanced time.

We first need to improve the
a-priori-bound from Lemma~\ref{apri2} under the assumption that
$a \geq 0$.
By Lemma~\ref{ingoing} and (\ref{p^1eq}),
\[
\frac{1}{2} a p_1 - \frac{1}{2} \frac{1+L/r^2}{p_1} = p^1 \leq - P
\]
on $\supp f_\mathrm{out}(v)$ for all $v>0$; $P>0$ by (\ref{suppcond}).
The fact that $a \geq 0$
and the assumptions on the support of the initial data imply that
\be \label{p_1est}
p_1 \leq \frac{1+L_+/r_0^2}{2P} \leq 1 \leq \frac{1+L/r^2}{p_1}
\leq \frac{1+L/r^2}{p_1} + a p_1
\ee
on $\supp f_\mathrm{out}(v)$. Since $a \geq 0$,
(\ref{p_1est}), (\ref{mform2}), and (\ref{mapri}) imply that
\be \label{goodt11est}
\int_{r_0}^\infty r^2 T_{11}(v,r)\, dr
\leq \int_{r_0}^\infty r^2 \left(S + a T_{11}\right)(v,r)\, dr
\leq \frac{1}{2 \pi} M.
\ee
Along any characteristic in $\supp f_\mathrm{out}$ and for $0<v<V$,
since $a \geq 0$,
\[
\left|\frac{dr}{dv}\right|
=
\frac{b}{2}\left(\frac{1+L/r^2}{(p_1)^2}-a \right)
\leq
\frac{1}{2} \left(1+\frac{L_+}{r_0^2}\right)\frac{1}{(p_1)^2}.
\]
Now we use the estimate (\ref{1/p1est}), but we control the $T_{11}$
integral in (\ref{-C1est}) by (\ref{goodt11est}) instead of Lemma~\ref{apri2}.
Hence
\[
\left|\frac{dr}{dv}\right|
\leq
\frac{1}{2} \left(1+\frac{L_+}{r_0^2}\right)\frac{1}{(p_-)^2}
\exp\left(\frac{2 M}{r_0^2} V + \frac{4 M}{r_0}\right).
\]
The assumption (\ref{nointerference}) now implies that $r>r_0$ for all
particles in $\supp f_\mathrm{out}(v)$ and $0<v\leq V$.
Hence by Corollary~\ref{Vcor} the solution exists on the interval
$[0,V]$.

We now show that at $v=V$ all the matter must be strictly
within $r< 2 M$
so that $a(V,r)< 0$ for some $r< 2 M$, which is a contradiction.
First we note that by (\ref{goodt11est}),
\beas
b(v,r)
&=&
\exp\left(- 4 \pi \int_r^\infty \eta T_{11}(v,\eta)\, d\eta\right)\\
&\geq&
\exp\left(- \frac{4 \pi}{r_0} \int_r^\infty \eta^2 T_{11}(v,\eta)\, d\eta\right)
\geq \exp\left(- \frac{2 M}{r_0}\right).
\eeas
Hence along any characteristic in $\supp f_\mathrm{out}$ by (\ref{p_1est}),
\[
\left|\frac{dr}{dv}\right|
=-\frac{p^1}{p^0}= b \frac{- p^1}{p_1} \geq
\exp\left(- \frac{2 M}{r_0}\right) \frac{P}{p_1}
\geq
\exp\left(- \frac{2 M}{r_0}\right) \frac{2 P^2}{1+L_+/r_0^2}.
\]
Assumption (\ref{tsform}) now implies that by $v=V$
all characteristics starting in $\supp \fn_\mathrm{out}$ are strictly
within $r< 2 M$, and the proof is complete.
\prfe
\section{After the formation of a trapped surface}
\label{afterts}
\setcounter{equation}{0}
In this section we investigate the Einstein-Vlasov system
(\ref{boundc})--(\ref{T22}) for data which are such that
$a(0,r_0) < 0$ for some $r_0>0$, i.e., the data already
contain a trapped surface. Besides being of interest in itself
this problem is relevant in connection with the result of the
previous section, where it was shown that such data evolve out of
data not containing a trapped surface, but where the argument stopped
after $a$ becomes negative somewhere;
notice that Theorem~\ref{ts} was proven by contradiction.
If we for example want to show that
eventually all matter ends up in the region $\{ r < 2 M\}$
and that $r= 2 M$ is complete and the event horizon of the evolving
black hole we have to be able to continue the analysis after
$a$ has become negative somewhere.

First we check that the system is well posed on the domain
$\{ r \geq r_0\}$. If $f$ and therefore $T_{11}$ and $S$ are
given for $r\geq r_0$, we define $b$ by (\ref{bformula}) as before.
Next we recall (\ref{alpharrewr}).
The desired boundary condition for $r a - r$ at infinity is
$\lim_{r\to \infty}(r a - r) = - 2 M$, and hence
\be \label{alphadef}
a(v,r) = 1 - \frac{2 m(v,r)}{r},
\ee
where
\be \label{moutdef}
m(v,r) = M - \frac{1}{2} \int_r^\infty 4 \pi \eta^2 (T_{11} + S)
\exp\left(-\int_\eta^r 4 \pi \sigma T_{11} d\sigma\right)\, d\eta
\ee
and $M>0$ denotes the ADM mass of the spacetime.

If $a(0,r_0) < 0$ for some $r_0>0$ then by Proposition~\ref{aneg},
$a(v,r_0) < 0$ for
all $v>0$ for which the solution exists.
Since characteristics can only leave but never enter the region
$\{ r \geq r_0\}$ when followed forward in $v$,
\[
f(v,r,p_1,L) = \fn ((R,P_1,L)(0,v,r,p_1,L))
\]
defines the solution of the Vlasov equation (\ref{vlasov}) on
$\{ r \geq r_0\}$ with initial data $\fn$.

We now specify the data which we consider in the
present section: $\fn \in C^1([r_0,\infty[ \times ]0,\infty[\times [0,\infty[)$
is non-negative and compactly supported in
$[r_0,\infty[ \times ]0,\infty[ \times [0,\infty[$. Defining
\[
\open{m}_\mathrm{\,out} :=
2 \pi \int_{r_0}^\infty  \eta^2 (\open{T}_{11} + \open{S}\,)
\exp\left(-\int_\eta^{r_0} 4 \pi \sigma \open{T}_{11} d\sigma\right)\, d\eta
\]
we choose some constant
\[
M > \open{m}_\mathrm{\,out} + \frac{r_0}{2}
\]
which plays the role of the ADM mass in (\ref{moutdef}). Then
\[
a(0,r_0) = 1 - \frac{2 m(0,r_0)}{r_0} < 0
\]
as desired. Notice that the system can be studied on
$\{r \geq r_0\}$ regardless of what represents the mass
in the central region $\{ r < r_0\}$ as long as that
mass is there initially. In particular, this
has the advantage that we need not take care that the outer
matter does not interfere with a central steady state
as was necessary in Theorem~\ref{ts}.
First, we obtain the following global existence result.
\begin{theorem} \label{glexout}
Initial data $\fn$ as specified above launch a unique
solution
$f\in C^1([0,\infty[ \times [r_0,\infty[ \times ]0,\infty[\times [0,\infty[)$
of the Einstein-Vlasov system (\ref{boundc})--(\ref{T22}).
\end{theorem}
\prf
As a first step one can establish a local existence result
corresponding to Theorem~\ref{locex}, using the arguments indicated
there and in the well-posedness discussion above.
If $V>0$ is the length of the maximal existence interval and
\[
\sup\left\{ p_1 + \frac{1}{p_1} \mid (r,p_1,L) \in \supp f(v),\ 0 \leq v < V
\right\} < \infty
\]
then $V=\infty$. But the required lower and upper bounds on $p_1$
along characteristics follow by the estimates in the proof of
Theorem~\ref{pbounds}.
\prfe
Let us now define $m_0:=M-\open{m}_\mathrm{\,out}>0$ as the
mass initially in the region $]0,r_0]$, and assume that
\[
\supp \fn \subset [r_0,R_1] \times [p_- ,p_+] \times [0,L_+]
\]
where
\[
R_1 > r_0,\ 0 < p_- < p_+,\ L_+ := 12 m_0^2,
\]
and we require that all particles move inward initially,
more precisely,
\be \label{Pcondrelaxed}
2 P > 1 + \frac{L_+}{(2 M)^2},
\ee
where as before,
\[
P := - \max \left\{p^1 \mid (r,p_1,L) \in \supp \fn \right\}.
\]
We obtain the following asymptotic behavior for
large advanced time.
\begin{theorem} \label{asympt}
For data $\fn$ as specified above
all particles
in the region $\{r \geq r_0\}$ are moving inwards,
$p^1 < -P$. Moreover, there exist $V^\ast >0$
and $r_0 < R^\ast < 2 M$ such that for $v \geq V^\ast$
all the matter is in the region $\{r < R^\ast\}$,
and $a(v,2 M) = 0$. The line $v\geq V^\ast,\ r= 2 M$
is a radially outgoing null geodesic which is future complete
and is the generator of the event horizon of the spacetime
for $v\geq V^\ast$.
\end{theorem}
\prf
First we observe that we can apply Lemma~\ref{ingoing}
also in the present situation on the region $\{r \geq r_0\}$;
notice that in the proof of that lemma no assumption
on the sign of $a$ was made. Hence $p^1 < -P$
for all the particles in that region. This implies that
\[
\left| \frac{dr}{dv} \right| = - b \frac{p^1}{p_1} \geq b \frac{P}{p_1}.
\]
for these particles. As long as $r\geq 2 M$ and hence $a \geq 0$
it follows that $b\geq\exp(-2 M/2 M) = 1/e$;
for this estimate we again rely on (\ref{goodt11est})
so we need the estimate corresponding to (\ref{p_1est}), but
now only for $r\geq 2 M$ which accounts for the relaxed condition
(\ref{Pcondrelaxed}).
By (\ref{p^1eq}) and since $a\geq 0$ for $r\geq 2 M$,
\[
\frac{1+L/r^2}{p_1} = - 2 p^1 + a p_1 \geq 2 P
\]
so that
\[
\left| \frac{dr}{dv}\right| \geq \frac{2 P^2}{e (1+L/r^2)}\geq
\frac{2 P^2}{e(1+L_+/(2 M)^2)}
\]
as long as  $r\geq 2 M$.
This shows that there exists $V^1 >0$ such that
for $v\geq V_1$ all particles
are in the region $\{r\leq 2 M\}$. If we choose some $r_0 < R < 2 M$
and $V^\ast:=V_1 + 1$, then
\[
c := \inf\left\{\frac{b(v,r)}{p_1} \mid
V_1 \leq v \leq V^\ast,\ R \leq r \leq 2 M,\
(r,p_1,L) \in \supp f(v) \right\} > 0.
\]
This implies that within the interval $[V_1,V^\ast]$ all
particles must have moved by a uniform distance to the left
which proves the existence of $R^\ast$.

For $v \geq V^\ast$ and $r\geq R^\ast$,
\[
a(v,r) = 1 - 2 M/r,\ b(v,r) = 1.
\]
The line $r=2M$, $v\geq V^\ast$ is null since $a(v,2 M)=0$,
and examining the geodesic equation for this
null geodesic shows
that it exists on an interval of affine parameter which is unbounded
to the right. The proof is complete.
\prfe

If we go back to Theorem~\ref{ts} we see that for a solution
as considered in that theorem there exists $v>0$ such that
$f_\mathrm{out}(v)$ defines data which
satisfy the assumptions of Theorem~\ref{asympt}. In particular,
there exists $0<r_0<2M$ such that $a(v,r_0)<0$,
and we obtain the following corollary.
\begin{cor} \label{corasympt}
The solutions obtained in Theorem~\ref{ts} exhibit the same
asymptotic properties as obtained in Theorem~\ref{asympt}.
\end{cor}

To conclude this paper we exploit our estimates
to construct a class of solutions where the initial data
represent a black hole surrounded by a shell of Vlasov matter.
These solutions will illustrate the fact that event horizons and
apparent horizons do in general not coincide.

\smallskip

\noindent
{\bf Remark.}
The fact that in general the apparent horizon and the event horizon
need not coincide  is usually illustrated
by sending a shell of so-called null dust, i.e., a pressure-less
fluid of photons, into a
black hole, and the corresponding spacetimes are
known as Vaidya spacetimes. The theorem below
shows that the corresponding behavior
of the horizons can also be achieved with a less
artificial matter model.

\smallskip

We consider
non-negative and compactly supported data
$\fn \in C^1 (]0,\infty[^2 \times [0,\infty[)$ with mass
\[
0 < \open{m}_\mathrm{\,out} = 2 \pi \int_0^\infty \eta^2 (\open{T}_{11}+\open{S}\,)
\exp\left(-\int_\eta^\infty 4 \pi \sigma \open{T}_{11}\, d\sigma\right)\,d\eta
< M
\]
and
\[
\supp \fn \subset [R_0,R_1] \times [p_-,p_+] \times [0,L_+]
\]
where
\[
0 < 2 M < R_0 < R_1,\ 0 < p_- < p_+,\
L_+ = 12 m_0^2 = 12 (M - m_\mathrm{out})^2.
\]
As before, we require that all particles are initially moving
inward sufficiently fast in the sense that (\ref{Pcondrelaxed})
holds.
For
$r^\ast := 2 m_0 < 2 M$ we have $a (0,r^\ast) =0$
while $a (0,r^\ast) > 0 / < 0$ for $r > r^\ast / r < r^\ast$.
In other words we have a trapped region $0 < r < r^\ast$, a black hole
of mass $m_0$, surrounded
by a shell of Vlasov matter which is moving inwards,
and we obtain the following result.
\begin{theorem} \label{vaidya}
Data as specified above launch a unique solution $f$
on the domain $[0,\infty[\times ]0,\infty[$.
All the particles in $\supp f$ are moving towards the center
which they all reach within a finite interval
$[0,\tau^\ast]$ of proper time. There exist $V^\ast > 0$
and $0 < R^\ast < 2 M$ such that $m(v,R^\ast) = M$ for $v\geq V^\ast$.
Hence if
\[
R(v) := \sup\{ r > 0 \mid a(v,r) < 0 \}
\]
then $R(v) = r^\ast$ on some interval of advanced time $[0,v^\ast]$
with $0 < v^\ast \leq V^\ast$,
and $R(v) = 2 M$ for $v\geq V^\ast$. On the other hand,
the generator of the event horizon is a
radially outgoing null geodesic which coincides
with  $R(v) = 2 M$ for $v\geq V^\ast$, but lies strictly to
the right of
$R(v) = r^\ast$ for $v\in [0,v^\ast]$.
\end{theorem}
\prf
If we take any $0 < r_0 < r^\ast$ we can apply Theorem~\ref{glexout}
to obtain a solution on $[0,\infty[ \times ]r_0,\infty[$.
If we decrease $r_0$ we get an extension of this solution,
and since  $0 < r_0 < r^\ast$ can be arbitrary the have the solution
on the asserted domain $[0,\infty[\times ]0,\infty[$.

Next we apply Lemma~\ref{ingoing}
which shows that all particles continue to move
inwards with $\sup \{p^1 \mid (v,r,p_1,L) \in \supp f \} < 0$
which proves the assertion on their behavior in proper time.
That all particles end up strictly
inside $\{r < 2 M\}$ within a finite interval
$[0,V^\ast]$ of advanced time can be shown exactly as above,
and the assertions
on the event horizon and the apparent horizon follow.
\prfe

\smallskip

\noindent
{\bf Concluding Remark.}
The analysis in the present paper leaves open the question
of what happens at the center $r=0$. Numerical evidence suggests
that in the situation of Theorem~\ref{ts} respectively
Theorem~\ref{asympt} a spacetime singularity arises where
the Kretschmann scalar
$R^{\alpha \beta \gamma \delta} R_{\alpha \beta \gamma \delta}$
blows up at $r=0$. At the same time the region where
$a < 0$ extends all the way to $r=0$ so that no causal
curve can enter $\{r>0\}$ out of the singularity
and strong cosmic censorship holds.
It remains to prove these assertions.

\end{document}